\documentstyle[epsf,epsfig,aps]{revtex}
\begin{document}
\preprint{\vbox{\hbox{DOE/ER/40762-180}\hbox{UMD PP\#99-106}}}

\title{Enhanced Signatures for Disoriented Chiral Condensates}
\author{Chi-Keung Chow and Thomas D.~Cohen}
\address{Department of Physics, University of~Maryland, College~Park, 
MD~20742-4111}
\date{\today}
\maketitle
\begin{abstract}
The probability distribution in $R$, the proportion of neutral pions to all 
pions emitted in heavy ion collisions in certain kinematic regions, has been 
suggested as a signal of a disoriented chiral condensate (D$\chi$C).  
Here we note that the signature can be greatly enhanced by making suitable 
cuts in the data.  
In particular, we consider reducing the data set such that the $k$ pions with 
lowest $p_T$ are all neutral. 
We find that, given such cuts, $\langle R \rangle$ can be substantially 
different from $1/3$.  
For example, for a single D$\chi$C domain without contamination due to 
incoherently emitted pions, $\langle R \rangle$ is $3/5$ given the pion with 
lowest $p_T$ is neutral, and $5/7$ given the two pions with lowest $p_T$ 
are both neutral, {\it etc.}. 
The effects of multi-domain D$\chi$C formation and noise due to incoherent 
pion emission can be systematically incorporated.  
Potential applications to experiments and their limitations are briefly 
discussed.  
\end{abstract}
\section{introduction}

Recently, there has been growing interest about the possibility of the 
formation of a disoriented chiral condensate (D$\chi$C) in heavy ion 
collisions \cite{And,KK,An1,An2,BK,Bj1,Bj2,RW1,KT,RW2,BKT1,BKT2,GGP,GM,CBNJ,CKMP,AHW,R,C,Bj3,old}.  
In an ultrarelativistic heavy ion collision, some region may thermalize at a 
temperature high enough so that chiral symmetry is restored in the region.  
As the system cools sufficiently rapidly back through the transition 
temperature, the chiral restored state is unstable as small fluctuations in 
any chiral direction ($\sigma,\vec{\pi}$) will grow exponentially.  
This can create regions where the pion field has a macroscopic occupation 
number.
It should be stressed that this scenario is not derivable directly from the
underlying theory of QCD and contains a number of untested dynamical
assumptions, principally that the cooling is rapid.  
While the failure of the system to form a D$\chi$C cannot be used to rule out 
that the system has reached the chiral restoration temperature (as the 
scenario described above is not derivable directly from QCD), observation of 
D$\chi$C formation would be a clear signal for chiral restoration at high 
temperature.  

There are clear signatures of D$\chi$C formation provided a single large 
domain is formed, containing a large number of pions.  
For example, one expects an excess in low $p_T$ pion production as the 
characteristic momentum of a pion from a large region is small. 
Such a signal works even if multiple regions of D$\chi$C are formed provided 
each region is large, but it is not decisive since one could imagine some 
other collective low energy effects which produce low $p_T$ pions.    
On the other hand, since the pions formed in a D$\chi$C, being essentially 
classical, form a coherent state, this coherent state has some orientation in 
isospace, and all of the pions in the domain are essentially maximally aligned 
(given the constraints of quantum mechanics) and point in the same isospin 
direction.  
If there are a large number of pions in the domain, this implies a 
distinctive distribution of $R$, the ratio of neutral to total pions in the 
domain \cite{And,KK,An1,An2,Bj1,Bj2,RW1,R}:  
\begin{equation}
f_0(R) \, =  \, \frac{1}{2 \sqrt{R}}.
\label{PofR}
\end{equation}
In contrast, the distribution from uncorrelated emissions is narrowly peaked 
about $1/3$ with a variance, $\langle R^2 \rangle - \langle R \rangle^2 = 
\frac{2}{9{\cal N}}$ where $\cal N$ is the total number of pions, and 
approaches a delta function at $R=1/3$ when ${\cal N}\to\infty$.   
Since these two distributions are dramatically different, this provides a  
clear signature for single domain D$\chi$C formation, provided one can 
kinematically separate the pions from the D$\chi$C from other pions in the 
system.  

Unfortunately, this signature depends critically on the assumption that a 
single large domain of D$\chi$C is formed, which is {\it a priori} rather 
unlikely.  
One expects the domains are of characteristic size $\tau$, the 
exponential growth time of the pion fluctuation in the unstable chiral 
restored state\cite{GGP,GM}.  
Since the size of the fireball is much larger than typical QCD length scales, 
it seems unlikely that $\tau$ would be of the size of the fireball, and thus  
formation of a single D$\chi$C domain is improbable.   
Formation of multiple domains of D$\chi$C, however, tends to wash out the 
$R$ distribution described by Eq.~(\ref{PofR}).  
As the pions emerging from different domains cannot be distinguished 
kinematically, by the central limit theorem the $R$ distribution will 
approach a normal distribution peaked at $R=1/3$.  
This normal distribution may be distinguished from the normal distribution 
arising from uncorrelated emission; the case of multiple domains of D$\chi$C 
will have a substantially larger variance.

Unfortunately, there is an important practical limitation which makes it 
difficult to exploit the $R$ distribution as a signature.   
Even under the most optimistic of scenarios, the total number of pions coming 
from D$\chi$C's will be a small fraction of the total number of pions.  
If one includes all pions produced in the reaction, the signal due to the 
pions from the D$\chi$C will presumably be overwhelmed.  
One can apply a low $p_T$ cut to suppress the noise due to incoherently 
emitted pions.  
However, even with the low $p_T$ cut, the noise may still be severe, 
as both the signal and the noise peak at $R=1/3$.  
In this paper, we suggest cuts which may dramatically enhance the 
signal-to-noise ratio.  
We study the conditional probability distribution of $R$ given only for the 
events in which the $k$ pions with the lowest $p_T$ are all neutral, and we 
will show that the expectation value of $R$ is shifted away from $1/3$.  
Since incoherent emission will result in a very narrow peak around $R=1/3$, 
any such shifts should be easily observable.  
Moreover, one can make successive cuts by increasing the value of $k$, and 
enhance the signal in each successive step. 

This paper is organized as follows.  
We will start with the simple scenario and study single domain D$\chi$C 
formation in Sec.~II.  
In Sec.~III the effects of multi-domain formation and the noises due to 
incoherently emitted pions will be studied, while potential experimental 
application and limitations will be discussed in Sec.~IV.  

\section{domain with non-zero isospin}

We are going to start with an unrealistic simple scenario and make it more 
realistic later on in our discussion.  
We will consider a single domain which is described by an isosinglet density 
matrix.  
(In general, a D$\chi$C is not a pure state; the wavefunction of the pion 
coherent state at the core of the ``fireball'' of the heavy ion collision 
is entangled with the energetic emission at the edge of the ``fireball''.) 
Such a state can be written as 
\begin{equation}
\rho = \sum_{n,I,I_z} c_{nI}{(-1)^{I_z}\over(2I+1)}\, 
|n,I,I_z\rangle \langle n,I,I_z|,  
\label{domain} 
\end{equation}
with $\sum_{n,I} c_{nI} = 1$.  
Note that the coefficients $c_{nI}$ are real, positive and do not depend on 
$I_z$ (as the full state is assumed to be isoscalar).  
The probability distribution of such a mixed state in the isospace is \footnote
{Here we are assuming that the typical $I$ is much less than 
$\langle n \rangle$.}
\begin{equation}
d^2P(\theta,\phi) = \sum_{n,I} c_{nI}/(2I+1) \, \sum_{I_z} 
|Y_{II_z}(\theta,\phi)|^2 \; \sin\theta d\theta\, d\phi 
= \sin\theta d\theta\, d\phi / 4\pi, \qquad 
dP(\theta) = \sin\theta d\theta/2, 
\end{equation}
where the angles $(\theta,\phi)$ are defined such that a unit vector in 
isospace is $(r_x,r_y,r_0) = (\sin\theta \cos\phi, \sin\theta \sin\phi, 
\cos\theta)$.  
The probability distribution is uniform, as the condensate is equally likely 
to point at any direction on the two-dimensional sphere, as demanded by 
isospin symmetry.  

The number operator of neutral pions in the condensate is given by 
\begin{equation}
n_0 = a_0^\dag a_0 = \vec a^\dag \cdot \vec a \cos^2\theta = \langle n \rangle 
\cos^2 \theta,  
\end{equation}
with $\vec a = (a_x,a_y,a_0)$ is a vector of hermitian annihilation operators 
which annihilate $\pi_x=(\pi_+ +\pi_-)/\sqrt{2}$, 
$\pi_y=(\pi_+ - \pi_-)/\sqrt{2} i$, and $\pi_0$, respectively, and 
$\langle n \rangle = \sum_{n,I} c_{nI} n$ is the expected number of pions.  
Since $R=n_0/\langle n \rangle$ \footnote{
We are assuming that $\langle n \rangle \gg 1$.}, 
one can easily calculate the probability distribution of $R$. 
\begin{equation}
dP = \textstyle{1\over2}\sin\theta d\theta = \textstyle{1\over2} d\cos\theta 
= d\sqrt{R} = \textstyle{1\over2} R^{-1/2} dR.   
\end{equation}  
By defining $dP \equiv f_0(R) \, dR$ (the subscript ``0'' stands for $I=0$), 
one has, in the limit that $\langle n \rangle$ is large, 
\begin{equation}
f_0(R) = \textstyle{1\over2} R^{-1/2},  
\end{equation}
recovering Eq.~(\ref{PofR}).  
The distribution is plotted in Fig.~1a.  
It is obvious that the shape is qualitatively different from the 
Poisson--Gaussian distribution due to incoherent emissions.  
The expectation value of $R$ is $1/3$,  
\begin{equation}
\langle R \rangle_0 \equiv \int_0^1 R f_0(R) dR = 1/3, 
\label{1/3}
\end{equation}
which has the simple interpretation that it is equally likely for the pion 
to be a $\pi_0$, $\pi_+$ or $\pi_-$, and hence on average a third of the 
pions are neutral.  

This distribution is a consequence of the fact that we have assumed the 
D$\chi$C to be an isosinglet, a reasonable assumption on physical grounds.  
However, let's consider the distribution of $R$ after the D$\chi$C emits a 
single $\pi_0$.  
The density matrix after the emission, which can be written as $\lambda \; a_0 
\, \rho \, a_0^\dag$, where $\lambda$ is a normalization constant, $\rho$ is 
the density matrix defined in Eq.~(\ref{domain}) and $a_0$ annihilates a 
neutral pion.  
This new density matrix is not an isosinglet.  
It is straightforward to show that the probability distribution for this 
state is 
\begin{equation}
dP = {\textstyle{1\over2}\sin\theta \cos^2\theta d\theta\over 
\int\textstyle{1\over2}\sin\theta \cos^2\theta d\theta} = 
{\textstyle{1\over2} R \cdot R^{-1/2} dR \over 
\int_0^1 \textstyle{1\over2} R \cdot R^{-1/2} dR}
\end{equation}
or equivalently, 
\begin{equation}
f_1(R) \equiv f(R|\hbox{1st pion is neutral}) 
= R f_0(R) \bigg/\int_0^1 R f_0(R) dR = \textstyle{3\over2} R^{1/2}.
\end{equation}
The distribution $f_1(R)$ is plotted in Fig.~1b, which is 
drastically different from $f_0(R)$.  
The distribution is skewed towards the high end, while $f_0(R)$ is skewed 
towards the low end.  
Moreover, the expectation value of $R$ is clearly pushed up:  
\begin{equation}
\langle R \rangle_1 \equiv \int_0^1 R f_1(R) dR = 3/5.  
\label{3/5}
\end{equation}
So we have arrived at the intriguing conclusion that, if the ``first pion'' 
emitted from a isosinglet D$\chi$C is neutral, 60\% of the pions 
subsequently emitted from the D$\chi$C are neutral, a huge enhancement from 
the original expectation of 33\%.  

\bigskip

\begin{figure}
\epsfig{file=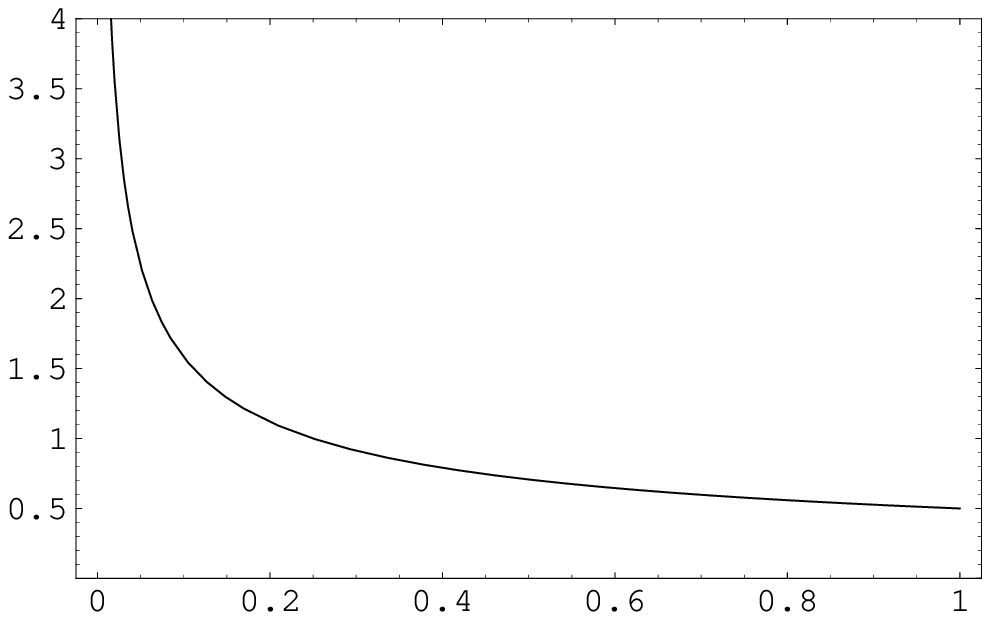, width=3.4in} 
\epsfig{file=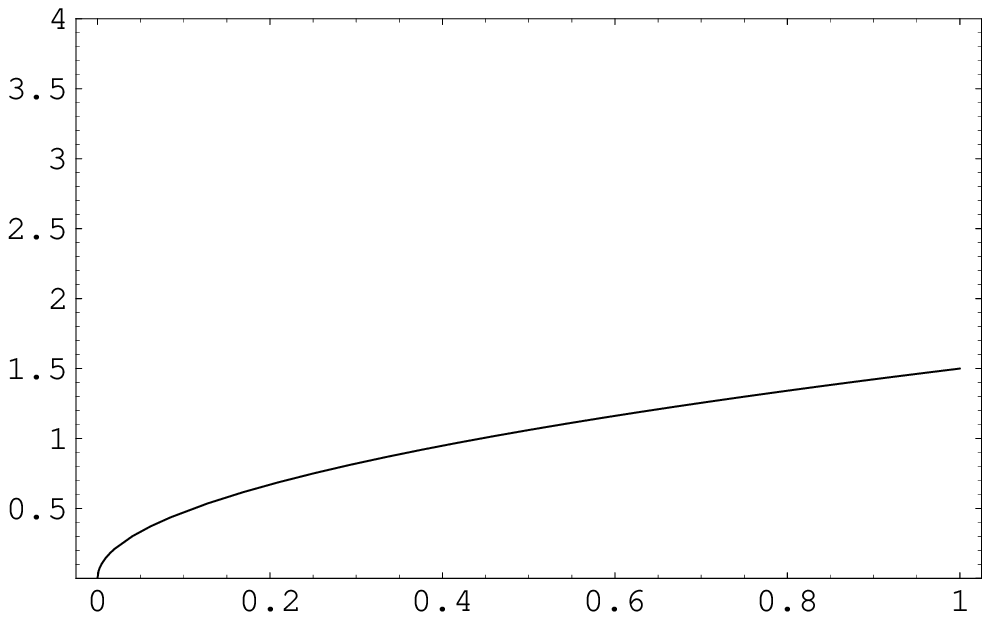, width=3.4in} 

\hskip 115pt Fig.~1a \hskip 220pt Fig.~1b 

\bigskip

\epsfig{file=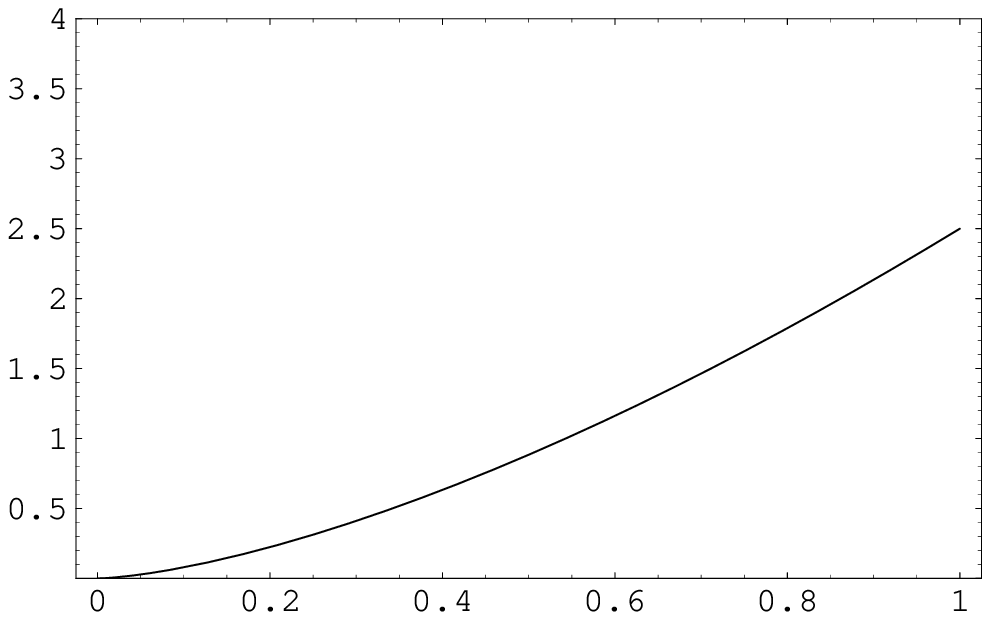, width=3.4in}
\epsfig{file=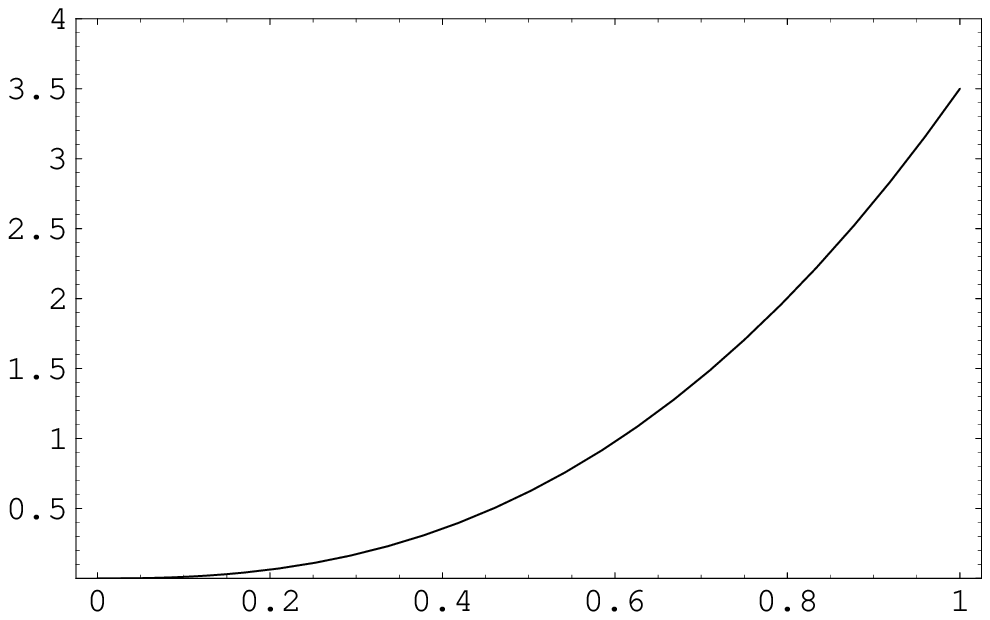, width=3.4in} 

\hskip 115pt Fig.~1c \hskip 220pt Fig.~1d

\bigskip

\caption{The probability distribution functions $f_k(R)$ for different values 
of $k$.
Fig.~1a, b, c and d are $f_0(R)$, $f_1(R)$, $f_2(R)$ and $f_3(R)$, 
respectively.  }
\end{figure}

\bigskip

This extraordinary statement certainly deserves more discussion.  
First, what is our criterion to decide which is the ``first pion''?  
The answer is simple: it can be any criterion.
It does not matter as long as it is {\it a priori\/} equally likely to be a 
$\pi_0$, a $\pi_+$ or a $\pi_-$.  
The derivation just depends on our removing a neutral pion from the isosinglet 
D$\chi$C.  
It can be the first pion emitted in time, or the last one emitted in time, or 
even the 17th emitted in time.  
The criterion can also be unrelated to the order of emission.  
For example, we can choose the ``first pion'' to be the one with the smallest 
polar angle.  
One can use any of these criteria to identify the ``first pion'', and if 
it turns out to be neutral, then the $R$ distribution of subsequent emissions 
is always given by $f_1(R)$, provided we are in the large number limit.  
However, this is only true in this idealized scenario, when all the pions 
are coming from a simple D$\chi$C domain.  
In reality, some of the pions come from incoherent emission, and if the 
``first pion'' turns out to be incoherently emitted, the expectation value 
of $R$ of the remaining pions is still going to be $1/3$, not $3/5$.  
As a result, we want to choose our criterion in such a way that the ``first 
pion'' is likely to originate from the D$\chi$C and not from incoherent 
emissions.  
Since D$\chi$C pions by hypothesis have low $p_T \sim 1/L$, where $L$ is the 
size of the domain, a natural choice is to use the pion with the lowest $p_T$ 
as our ``first pion''.  

After clarifying the meaning of the term ``first pion'', we move on to 
discuss the physical origin of the modification of the probability 
distribution of $R$.  
In a nutshell, we are seeing the physics of (iso)spin alignment due to 
Bose condensation.  
To illustrate the point, let us first consider the following apparently 
unrelated Stern--Gerlach experiment.  
Consider a large number of massive spin-1 particles, which for concreteness 
will be called deuterons.  
Initially they are all polarized along a randomly chosen direction 
$\vec n$, which is {\it a priori\/} equally likely to be any direction 
in three dimensional space.  
In other words, $\vec S \cdot \vec n = 0$ for all the deuterons.  
Now let us pick one of these deuterons and pass it through a Stern--Gerlach 
spectrometer which measures $S_z$, the spin along the $z$-axis.  
What is the probability that the measurement gives $S_z=0$?  
The answer is clearly $1/3$, as the cases for $S_z=+1$, 0 and $-1$ are 
equally likely.  
On the other hand, if the measurement on the first deuteron gives $S_z=0$, 
what is the conditional probability for the next deuteron to pass through 
the Stern--Gerlach spectrometer also to be measured to have $S_z=0$?  
The answer this time is no longer $1/3$.  
The spins of all the deuterons are aligned along the same direction $\vec n$, 
and that the first deuteron is measured to have $S_z=0$ suggests $\vec n$ 
is more probable to be more or less aligned along $\vec z$ than otherwise.  
As a result, the conditional probability is no longer $1/3$, but can be 
easily shown to be $3/5$, which is exactly the predicted value for $\langle 
R \rangle_1$ in Eq.~(\ref{3/5}).  
The situation for a single domain of D$\chi$C is analogous, with 
isospin aligned pions instead of spin aligned deuterons.  
By construction, the pions in a D$\chi$C domain are isospin aligned, and by 
the same analysis, we have shown that the knowledge of the ``first pion'' 
being neutral can dramatically modify the conditional probability 
distribution of $R$.  

One can also consider the conditional probability distribution of $R$ 
in the case that the ``first pion'' is charged.  
Note that
\begin{equation}
f_0(R) = \textstyle{1\over3}\big(f(R|\hbox{1st pion is a $\pi_+$}) + 
f(R|\hbox{1st pion is a $\pi_-$}) + f(R|\hbox{1st pion is a $\pi_0$})\big), 
\end{equation}
and hence, since $f_1(R) = f(R|\hbox{1st pion is a $\pi_0$})$, 
\begin{equation}
\tilde f(R) \equiv f(R|\hbox{1st pion is charged}) = \textstyle{3\over2} 
f_0(R) - \textstyle{1\over2} f_1(R) = \textstyle{3\over4} (1-R) R^{-1/2}.
\end{equation}
The expectation value of $R$, given that the ``first pion'' is charged, can be 
easily shown to be $1/5$.  
As a consistency check, one can calculate $\langle R \rangle$, the expectation 
value of $R$ regardless of the species of the ``first pion''.  
Since the ``first pion'' is twice as likely to be charged as to be neutral, 
\begin{equation}
\langle R \rangle_0 = \textstyle{1\over3}(\textstyle{3\over5}
+2 \times \textstyle{1\over5}) = \textstyle{1\over3}, 
\end{equation}
agreeing with Eq.~(\ref{1/3}).  

Lastly, we will study the conditional probability distribution of $R$ given 
that the $k$ pions with the lowest $p_T$, which will be hereafter referred to 
as the ``first $k$ pions'', are all neutral.  
It is straightforward to show that in this case 
\begin{equation}
dP = {\textstyle{1\over2}\sin\theta \cos^{2k}\theta d\theta\over 
\int\textstyle{1\over2}\sin\theta \cos^{2k}\theta d\theta} = 
{\textstyle{1\over2} R^k \cdot R^{-1/2} dR \over 
\int_0^1 \textstyle{1\over2} R^k \cdot R^{-1/2} dR}. 
\end{equation}
and 
\begin{equation}
f_k(R) \equiv f(R|\hbox{1st $k$ pions are all neutral}) 
= R^k f_0(R) \bigg/\int_0^1 R^k f_0(R) dR = (k+\textstyle{1\over2}) R^{k-1/2}.
\end{equation}
The distributions $f_2(R)$ and $f_3(R)$ are plotted in Fig.~1c and d,  
respectively.  
One can see that as $k$ increases, the distribution is more and more skewed 
towards the high end.  
As a result, the expectation value of $R$ increases with $k$.  
\begin{equation}
\langle R \rangle_k \equiv \int_0^1 R f_k(R) dR = (2k+1)/(2k+3).  
\label{R}
\end{equation}
It is useful to define $Q$ as the ratio of the number of $\pi_+$ 
to the number of total pions emitted.  
By symmetry it is also the ratio of the number of $\pi_-$ to the number of 
total pions emitted, and since $R+2Q=1$, 
\begin{equation} 
\langle Q \rangle_k = 1/(2k+3). 
\label{Q} 
\end{equation}

From the above analysis, the prescription to enhance the collective signal 
is quite clear.  
One should make successive cuts on the data sample on the condition that 
the $k$ pions with the lowest $p_T$ are all neutral, and measure $\langle 
R \rangle_k$ after each cut to see if it increases as predicted in 
Eq.~(\ref{R}).  
This result, however, depends on the assumption that we have only a single 
domain of D$\chi$C without any contamination due to incoherent pion 
emissions.  
Since this assumption is unrealistic for heavy ion collision experiments, 
the scenario we studied in this section is only an idealized situation.  
In the next section, we will discuss more realistic scenarios.  

\section{The effects of multi-domain formation and incoherent emissions}

The scenario considered in the last section is highly unrealistic in at least 
two ways.
First, as discussed in the introduction, single domain D$\chi$C formation is 
highly unlikely.  
For a realistic treatment one must study D$\chi$C formation with more 
than one domain, each pointing in a different direction in the isospace.  
Moreover, we have neglected the effect of incoherently emitted pions, which 
have very important effects.  
If the neutral ``first pion'' is incoherently emitted, the $R$ distribution 
of the remaining pions is described by $f_0(R)$, instead of $f_1(R)$ when 
the ``first pion'' comes from the D$\chi$C.  
In this section, we will incorporate these two effects and see how the 
predictions above are modified.  

We will study the expectation value of $R$, or equivalently the expectation of 
$Q$, for a situation described by the following parameters.  
The coherent fraction $\chi$ is the fraction of pions which originate from 
D$\chi$C domains, so that when $\chi=1$, all pions are coherently emitted, 
and when $\chi=0$, all pions are incoherently emitted.  
We will consider the case where there are $N$ domains, all containing an equal 
number of pions\footnote
{This assumption of all domains having the same number of pions is unrealistic 
but is made for illustrative purposes.  
The effects of unequal domain sizes will be briefly discussed below.}, which 
will be assumed to be large.    
Each domain is described by an isosinglet density matrix, but the isospins of 
pions in different domains are uncorrelated.    
Now the question is: if the ``first $k$ pions'' in this channel are all 
neutral, what are the expectation values of $R$ and $Q$ among the rest of the 
pions?  

The answer turns out to be the following expression:   
\begin{mathletters}
\begin{equation}
\langle R \rangle = \textstyle{1\over3} + 2\Delta, \qquad 
\langle Q \rangle = \textstyle{1\over3} - \Delta.
\end{equation}
The shift $\Delta$ is given by 
\begin{equation}
\Delta =  \chi \sum_{j=0}^k P_j \,  ({1\over 3} - {1\over 2j+3}) 
= \chi \big({1\over 3} - \sum_{j=0}^k P_j \,  {1\over 2j+3} \big),    
\end{equation}
where 
\begin{equation}
P_j = {k \choose j} \; p^j \, (1-p)^{k-j}, \qquad p=\chi/N. 
\end{equation}
\label{result}
\end{mathletters}
Each term in this formula has a simple interpretation: 

$\bullet$ The expectation value $\langle R \rangle$ is always $1/3$ for the 
incoherently emitted pions.  
Only the pions coming from the domains are affected by isospin alignment; 
hence the outstanding factor of $\chi$.  

$\bullet$ Each coherently emitted pion comes from one of the domains, which 
will be called domain X.  
How many of the ``first $k$ pions'' also come from domain X?  
The probability for each pion coming from domain X is $p=\chi/N$, and the 
probability that $j$ of the ``first $k$ pions'' coming from domain X is 
$P_j = {k \choose j} \, p^j \, (1-p)^{k-j}$.  

$\bullet$ Given that $j$ of the ``first $k$ pions'' is coming from domain X, 
the conditional expectation value of $Q$ decreases from $1/3$ to $1/(2j+3)$, 
while the conditional expectation value of $R$ increases by twice the above 
quantity.   

In passing, we note that $\Delta$ can also be expressed as an integral or 
the hypergeometric function $_2F_1$: 
\begin{eqnarray}
\Delta &=& \chi \Big({1\over 3} - {1\over \sqrt{p^3}} 
\int_0^{\sqrt{p}} dz \; z^2 (z^2 + 1 - p)^k \Big) \nonumber \\  
&=& \chi \Big({1\over 3} - {\cos^{2k+3}\Theta \over \sin^3\Theta} 
\int_0^\Theta d\vartheta\; {\sin^2\vartheta \over \cos^{2k+4}\vartheta} \Big), 
\qquad \tan^2\Theta = {\chi/N \over 1- \,\chi/N} \nonumber \\
&=& {1\over 3} \chi \Big(1-(1-p)^k \, 
_2F_1[{3\over2},-k,{5\over2},{-p\over 1-p}] \Big).  
\end{eqnarray}
In Fig.~2, we have made contour plots of $\Delta=1/30$ (such that $\langle R 
\rangle = 0.4$ and $\langle Q \rangle = 0.3$), for $k=1,\dots,5$ in the 
$(\chi,1/N)$ parameter space.  
The horizontal axis is the coherent fraction $\chi$ while the vertical axis is 
$1/N$ where $N$ is the number of domains.  
Both $\chi$ and $1/N$ range from 0 to 1.  
Thus, for example, with 3 domains and $\chi=0.6$, in order to have $\Delta 
\geq 1/30$ we must have $k\geq 3$.  

Equations (\ref{result}) illustrate the main results of this paper.  
One can see that, without any D$\chi$C formation, $\chi=0$ (corresponding to 
the left edge of Fig.~2), $\Delta$ vanishes, and $\langle R \rangle = 
\langle Q \rangle = 1/3$ as expected.  
The bottom edge of the plot corresponds to $N\to\infty$ and also gives 
$\Delta = 0$ for any finite value of $k$.  
The shift $\Delta$ is largest for a single domain of D$\chi$C without 
any noise due to incoherently emitted pions, {\it i.e.}, when $\chi=N=1$ (the 
top right corner of the contour plots), giving $\Delta = 1/3 \, - \, 
1/(2k+3)$ and reproducing Eqs.~(\ref{R}) and (\ref{Q}).  
For fixed values of $(\chi,N)$, $\Delta$ increases with $k$, accounting for 
the spreading of the parameter space with $\Delta > 0.4$ as $k$ increases  
from 1 to 5 in Fig.~2.  

\bigskip

\begin{figure}
\hskip 120pt $1/N$

\centerline{ \epsfig{file=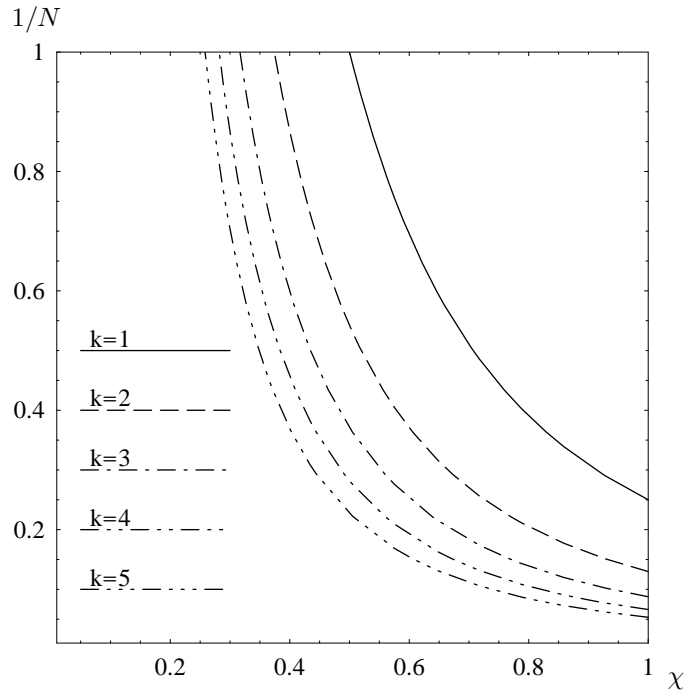, width=3.4in} $\chi$}

\bigskip

\caption{Contour plots of $\Delta = 1/30$ (such that $\langle R \rangle =
0.4$ and $\langle Q \rangle = 0.3$) for different values of $k$.  
The horizontal axis is the coherent fraction $\chi$, while the vertical axis 
is $1/N$ where $N$ is the number of domains.  
Both $\chi$ and $1/N$ range from 0 to 1. 
The curves, from top right to bottom left are for $k=1$, 2, 3 4 and 5,  
respectively. 
The shift $\Delta$ is larger than $0.4$ above the curves and smaller than 
$0.4$ below the curves. }
\end{figure}

\bigskip

One expects that when the number of D$\chi$C domains is large ($N\gg1$) or 
when most of the pions are incoherently emitted ($\chi\ll 1$), it will be 
difficult to observe clear signals of D$\chi$C formation.  
However, in such situations $\chi/N$ is small and $\Delta$ is dominated by 
the $j=1$ term (the $j=0$ term always identically vanishes) and 
\begin{equation}
\Delta = {2 \chi^2 k \over 15 N} + {\cal O}({\chi^3\over N^2}).  
\label{small}
\end{equation}
Thus a large $k$ may make up for a small coherent fraction $\chi$, or a large 
number of domains $N$, and enhance $\Delta$, which describes the shift of 
$\langle R \rangle$ and $\langle Q \rangle$ from $1/3$, to an experimentally 
measurable magnitude.  
From the form of Eq.~(\ref{small}), one expects this shift to be substantial 
whenever $k \sim N/\chi^2$. 
However, even for a value of $k$ as small as $N/4\chi^2$, $\Delta = 1/30 + 
{\cal O} (\chi^3/N^2)$, which translates to $\langle R \rangle = 0.4$ and 
$\langle Q \rangle = 0.3$ --- a substantial deviation from the incoherent 
case.  
This suggests one should make successive cuts for events where the $k$ pions 
with lowest $p_T$ are all neutral, and study $\langle R \rangle$ after each 
cut.  
An increase of $\langle R \rangle$ with $k$ would suggest that D$\chi$C 
domains are formed.  

Equation (\ref{small}) appears to suggest that one can increase $\Delta$ to 
an arbitrarily large magnitude by choosing a sufficiently large value of $k$. 
Of course this is not true.  
Equation (\ref{small}) is obtained as the leading term in a $\chi/N$ 
expansion, but when $k \to \infty$, this expansion breaks down as terms of 
higher order in $\chi/N$ are enhanced by factors of $k \choose j$.  
We can easily see that 
\begin{equation}
\Delta \to \textstyle{1\over3}\chi, \quad 
\langle R \rangle \to \textstyle{1\over3}+\textstyle{2\over3}\chi, \quad
\langle Q \rangle \to \textstyle{1\over3}(1-\chi), \qquad k\to\infty 
\hbox{ with $\chi$ and $N$ fixed.} 
\label{limit}
\end{equation}
In other words, our signal enhancement scheme is fundamentally limited by the 
amount of noise due to incoherently emitted pions.  
When $\chi$ is small, most of the pions are incoherently emitted, and for 
them, $\langle R \rangle$ is always around $1/3$ regardless of what cuts one 
makes.  
On the other hand, the large $k$ limit of $\langle R \rangle$ does not depend 
on $N$, the number of D$\chi$C domains.  
Recall that we have several distinctive signatures, like the $R$ distribution 
in Eq.~(\ref{PofR}) and the conditional expectation values for $R$ described 
in the previous section, for a single domain of D$\chi$C, where all the pions 
in the D$\chi$C are isospin aligned.  
With multi-domain formation, where the pions in different domains may point 
to different directions in isospace, the effect of isospin alignment is 
greatly washed out.  
However, given that the ``first $k$ pions'' are all neutral with $k\gg N$, 
it is probabilistically extremely likely that each of the $N$ domains is 
the origin of some of these ``first $k$ pions''.  
As a result, each of these $N$ domains are well-aligned along the $\pi_0$ 
direction, and hence also well-aligned with each other.  
As $k\to\infty$, the $N$ domains look more and more like a single big 
domain in the $\pi_0$ direction, which is the case where the signature is 
the most dramatic.  
In short, with large $k$, our cuts are picking out the events where the 
signals are the strongest, and hence resulting in a large signal-to-noise 
ratio.  

In above we have assumed the sizes of all $N$ domains are identical for 
illustrative purposes.  
A more realistic treatment would have $N$ domains, each with different 
sizes $p_i$, $1\leq i \leq N$, such that $\sum_i p_i = \chi$.  
(The size of a domain is defined to be the fraction of pions which 
originate from this particular domain.)  
Then Eqs.~(\ref{result}) are generalized to 
\begin{equation}
\Delta = {1\over 3} \chi - \sum_{i=1}^N p_i \; \sum_{j=0}^k {k \choose j} 
p_i^k (1-p)^{k-j} \; {1\over 2j+3}.  
\end{equation}
In the weak signal limit, {\it i.e.}, when all the $p_i$'s are small, 
Eq.~(\ref{small}) gets modified to 
\begin{equation}
\Delta = {2k \over 15} \sum_{i=1}^N p_i^2 + {\cal O}({\chi^3\over N^2}) 
= {2k \over 15} \overline p + {\cal O}({\chi^3\over N^2}) , 
\end{equation}
where $\overline p = \sum_i p_i^2$ has the following nice interpretation: 
$\overline p$ is the average over all pions (both coherently and incoherently 
emitted) of the sizes of the originating domains, which is $p_i$ for a pion 
from domain $i$ and zero for an incoherently emitted pion.   
For $N$ domains of equal sizes, $\overline p = \chi^2/N$ and Eq.~(\ref{small}) 
is recovered.  
Again we see that $\Delta$ grows linearly with $k$ in the weak signal limit.  
As $k\to\infty$ the shift $\Delta$ is again limited by the bound 
(\ref{limit}), which applies also for the cases of unequal domain sizes.  

\section{summary}

To recapitulate, we suggest the following experimental procedures:  

$\bullet$ Count the number of neutral and charged pions {\it event by event\/} 
from heavy ion collision experiments and measure their individual transverse 
momenta and rapidities.  

$\bullet$ Apply a low $p_T$ cut to suppress the noise due to uncorrelated 
pion emission. 

$\bullet$ Bin the events in different rapidity windows.  

$\bullet$ In each rapidity window, calculate the expectation value 
$\langle R \rangle$.  

$\bullet$ Make a cut to retain only events where the pion with the lowest 
$p_T$ is neutral.  

$\bullet$ Calculate, in each rapidity window, the expectation value 
$\langle R \rangle$ for all remaining pions in all events which survive the 
cut.  

$\bullet$ Make another cut on the surviving events to retain only those where 
the pion with the second lowest $p_T$ is also neutral.  

$\bullet$ Again, calculate in each rapidity window the expectation value 
$\langle R \rangle$ for all remaining pions in all events which survive the 
cuts. 

$\bullet$ Repeat the above prescription of making successive cuts to retain 
only events in which the pion with the next lowest $p_T$ is also neutral, 
and calculate $\langle R \rangle$ for each rapidity window after each cut.  
If we find $\langle R \rangle$ deviates from $1/3$ then we are seeing 
signatures from D$\chi$Cs.  

Note that this prescription requires reconstructions of $p_T$'s of 
individual pions, both charged and neutral.  
We have also presumed that the coherent fraction $\chi$ and the number of 
domains formed $N$ are roughly the same for each event.  
(More specifically, the probability distributions for $\chi$ and $N$ are 
narrow peaked.)  

By applying these successive cuts, we are retaining the events with D$\chi$C 
formation {\it and\/} most of the pions are well-aligned along the $\pi_0$ 
direction.  
What is being cut are the events with D$\chi$C formation but most of 
the pions are well-aligned along the $\pi_x$ or $\pi_y$ directions, and the 
events where there are incoherent pions with very low $p_T$, which is the main 
source of noise to our signal.  
As a result, these successive cuts are substantially improving the 
signal-to-noise ratio, making it easier to observe D$\chi$C formation.  
On the other hand, just like any other cuts on data to suppress the noises, 
we are giving up on statistics.  
Moreover, for large $k$ we are cutting on rare events so the loss in 
statistics can be severe.  
For the cases where the signal is weak (small coherent fraction $\chi\ll 1$ or 
large number of domains $N\gg 1$) on each cut we are losing about two-thirds 
of the events.  

In conclusion, we have devised new cuts to enhance the signal in searches for 
D$\chi$C.  
These cuts retain only events where the $k$ pions with lowest $p_T$ are 
all neutral.  
We have shown that, after these cuts, the fraction of neutral pions within 
the remaining sample is substantially larger if D$\chi$Cs are formed in the 
heavy ion collision.  

\bigskip

Support of this research by the U.S.~Department of Energy under grant 
DE-FG02-93ER-40762 is gratefully acknowledged.

\end{document}